\documentclass[aps,prb,twocolumn,amssymb,amsmath,superscriptaddress]{revtex4-1}
\usepackage{bm}
\usepackage{times}
\usepackage{graphicx}
\usepackage{color}
\usepackage{dcolumn}
\usepackage[colorlinks=true, letterpaper=true, pdfstartview=FitV, linkcolor=blue, citecolor=blue, urlcolor=blue]{hyperref}
\usepackage{appendix}
\usepackage[normalem]{ulem}

\begin{document}
\title{Transport features of topological corner states in honeycomb lattice with multihollow structure}
\author{Kai-Tong Wang}
\affiliation{School of Physics and Engineering, Henan University of Science and Technology, Luoyang 471023, China}
\affiliation{College of Physics and Optoelectronic Engineering, Shenzhen University, Shenzhen 518060, China}
\author{Fuming Xu}
\email[]{xufuming@szu.edu.cn}
\affiliation{College of Physics and Optoelectronic Engineering, Shenzhen University, Shenzhen 518060, China}
\author{Bin Wang}
\affiliation{College of Physics and Optoelectronic Engineering, Shenzhen University, Shenzhen 518060, China}
\author{Yunjin Yu}
\affiliation{College of Physics and Optoelectronic Engineering, Shenzhen University, Shenzhen 518060, China}
\author{Yadong Wei}
\affiliation{College of Physics and Optoelectronic Engineering, Shenzhen University, Shenzhen 518060, China}

\begin{abstract}
Higher-order topological phase in 2-dimensional (2D) systems is characterized by in-gap corner states, which are hard to detect and utilize. We numerically investigate transport properties of topological corner states in 2D honeycomb lattice, where the second-order topological phase is induced by an in-plane Zeeman field in the conventional Kane-Mele model. Through engineering multihollow structures with appropriate boundaries in honeycomb lattice, multiple corner states emerge, which greatly increases the probability to observe them. A typical two-probe setup is built to study the transport features of a diamond-shaped system with multihollow structures. Numerical results reveal the existence of global resonant states in bulk insulator, which corresponds to the resonant tunneling of multiple corner states and occupies the entire scattering region. Furthermore, based on the well separated energy levels of multiple corner states, a single-electron source is constructed.
\end{abstract}
\maketitle

\section{introduction}

Topological phases of matter have been the central issues of condensed matter study in the past decades\cite{XLQi,Ren16,Qi19,Niu20}, which is recently extended to higher-order band topology\cite{HOTI1}. Initially addressed as quantized electric multiple insulators\cite{HOTI2}, higher-order topological insulators (HOTIs) are described by new topological invariants\cite{HOTI3,HOTI4} and have different bulk-boundary correspondence. In 2-dimensional (2D) systems, the second order topological phases are characterized by 0D in-gap corner states, while a 3D HOTI may host 1D gapless hinge states or in-gap corner states\cite{HOTI1}. Up to now, HOTIs have been reported in various systems\cite{ingap,pijunction,Tao20NJP,Li20PRB,Jiang20SCIBULL,Cheng21PRB,Liu21PRB,Liu21PRL,wang21PRL}, including quasicrystals\cite{Chen20PRL,Hua20PRB,Huang22} and Anderson insulators\cite{Yang21PRB,zhang21PRL}.

Honeycomb lattice materials, such as graphene and MoS$_2$, are ideal platforms for exploring topological phases\cite{Qiao11,Xiao12,Jiang15FOP,Ren16,Xu16NJP,Xu17FOP,Xu17PRB}. The famous Kane-Mele model of quantum spin Hall effect is first theoretically proposed on grapehe\cite{Kane}, which is later demonstrated to have an extremely small band gap\cite{Yao1}. Honeycomb lattice materials supporting the Kane-Mele model are continuously found, for instance, silicene\cite{Yao2,Yao3,Ezawa,Qiao14}, germanium\cite{Yao3}, and stanene\cite{XuYong13,XuYong20,Zhao20}, etc. First-principles calculation suggested jacutingaite Pt$_2$HgSe$_3$ as a new Kane-Mele quantum spin Hall insulator\cite{Marzari1,Marzari2,Qiao21PRB}, where a 0.1 $eV$ band gap is achieved in experiment very recently\cite{Kandrai}. A modified Kane-Mele model with in-plane Zeeman field is shown to be second-order topological insulator\cite{Ren2020}, where transport induced dimer state is formed during the resonant tunneling of two topological corner states\cite{KTWang}. The topological corner states are embedded in a large energy gap and localized in real-space obtuse corners, which makes them hard to detect via transport means.

To increase the probability of observing these corner states, a possible solution is creating appropriate locations to host more corner states in one single system. In this work, based on the modified Kane-Mele model\cite{Ren2020,KTWang}, we propose a multihollow structure through precise engineering of 2D honeycomb lattice. The existence of multiple corner states is revealed in the energy spectrum as well as eigenfunction distributions, and their transport features are numerically investigated in a two-probe setup. Both the transmission function and partial local density of states confirm that, multiple corner states contribute to resonant tunneling and result in global resonant states in a large range of system parameters. We suggest a single-electron source via adiabatic pumping of multiple corner states, and the pumped charge is found to be well quantized.

The rest of the paper is organized as follows. In Sec.II, the model Hamiltonian for topological corner states in honeycomb lattice is introduced, and quantum transport formalism is briefly reviewed. Sec.III includes numerical results and relevant discussion on transport features of the multihollow structure. In addition, a single-electron source based on this multihollow system is proposed. A summary is finally given in Sec.IV.

\section{Model and Formalism}

We start with the modified Kane-Mele model in the presence of an in-plane Zeeman field, where second-order topological corner states reside in real-space obtuse intersections of the honeycomb lattice\cite{Ren2020}. The tight-binding Hamiltonian of this model is written as\cite{Ren2020}
\begin{align}
H_C = -t\sum_{<ij>} d_{i}^{\dag}d_{j} + it_{so}\sum_{\ll ij \gg} \nu_{ij}d_{i}^{\dag}s_{z}d_{j}
       + \lambda \sum_{i} d_{i}^{\dag} \mathbf {B}\cdot \mathbf {s}d_{i}. \label{Eq1}
\end{align}
Here $d_{i}^{\dag}=[d_{i,\uparrow}^{\dag},d_{i,\downarrow}^{\dag}]^{T}$ is the creation operator at site $i$ for both spin up ($\uparrow$) and down ($\downarrow$). $t$ in the first term is the nearest-neighbor hopping energy, which is taken as the energy unit. The second term denotes intrinsic spin-orbit coupling (SOC) with strength $t_{SO}$, which is represented by next-nearest-neighbor hopping. The first and the second terms together give the conventional Kane-Mele model for quantum spin Hall effect. $\nu_{ij}= \pm 1$ if the electron takes a clockwise or anticlockwise turn from site $j$ to $i$. The in-plane Zeeman field is introduced in the third term, with $\mathbf{B}=(B_x,B_y,0)$ and $\lambda$ the Zeeman field strength. This Zeeman field breaks the time-reversal symmetry and hence destroys quantum spin Hall effect. For a certain field strength, such as $\lambda \geq 0.2$\cite{Ren2020}, a pair of zero-energy corner states appear at the obtuse intersections of two different zigzag boundaries in a diamond-shaped honeycomb lattice, as shown in Fig.\ref{fig1}(a), which are the second-order topological phase protected by the mirror symmetry.

To study transport features of topological corner states, the diamond-shaped region hosting corner states is connected to two semi-infinite leads. The lead is chosen to be free of magnetic field, and its Hamiltonian follows the conventional Kane-Mele model:
\begin{align}
H_L = -t\sum_{<mn>} c_{m}^{\dag}c_{n} + it_{so}\sum_{\ll nm \gg} \nu_{mn}c_{m}^{\dag}s_{z}c_{n},
\end{align}
where $c_{m}^{\dag} / c_{m}$ is the creation/annihilation operator in this lead and $n/m$ labels its lattice site.

Influence of the $\alpha$ lead is treated as the self-energy, ${\Sigma}_{\alpha}^{r}$, which is conveniently evaluated through the transfer matrix approach\cite{Sancho}. As one of the most important transport properties, the transmission function of an electron with energy $E$ from lead $\beta$ to lead $\alpha$ is expressed in terms of the nonequilibrium Green's functions as\cite{Datta}
\begin{equation}\label{Trans}
T_{\alpha \beta} = \rm{Tr}[\Gamma_{\alpha} G^{r}(E) \Gamma_{\beta} G^{a}(E)],
\end{equation}
where $G^{r}$/$G^{a}$ is the retarded/advanced Green's function of the central region, $G^{r}(E) = (E-H_C-\sum_{\alpha=L,R}{\Sigma}_{\alpha}^{r})^{-1}$ and $G^{a} = G^{r\dag}$. The linewidth function of lead $\alpha$ is defined as $\Gamma_{\alpha} = i({\Sigma}_{\alpha}^{r}-{\Sigma}_{\alpha}^{a})$, and its eigenvector form can effectively speed up the calculation\cite{Guo09}.

The partial local density of states (PLDOS)\cite{PLDOS1,PLDOS2,PLDOS3} is another useful tool in transport study, which describes the real-space distribution of a scattering state:
\begin{equation}\label{Ldos}
dos_{\alpha}(i) = \frac{1}{2\pi} [G^{r} \Gamma_{\alpha} G^{a}]_{ii}.
\end{equation}
'Partial' means only electrons incident from lead $\alpha$ contribute to $dos_{\alpha}$, and $i$ represents the real-space lattice site. Transport properties including transmission and PLDOS of multiple corner states are numerically investigated and thoroughly discussed in the following section.

\section{NUMERICAL RESULTS AND DISCUSSION}

\begin{figure}[tbp]
\centering
\includegraphics[width=\columnwidth]{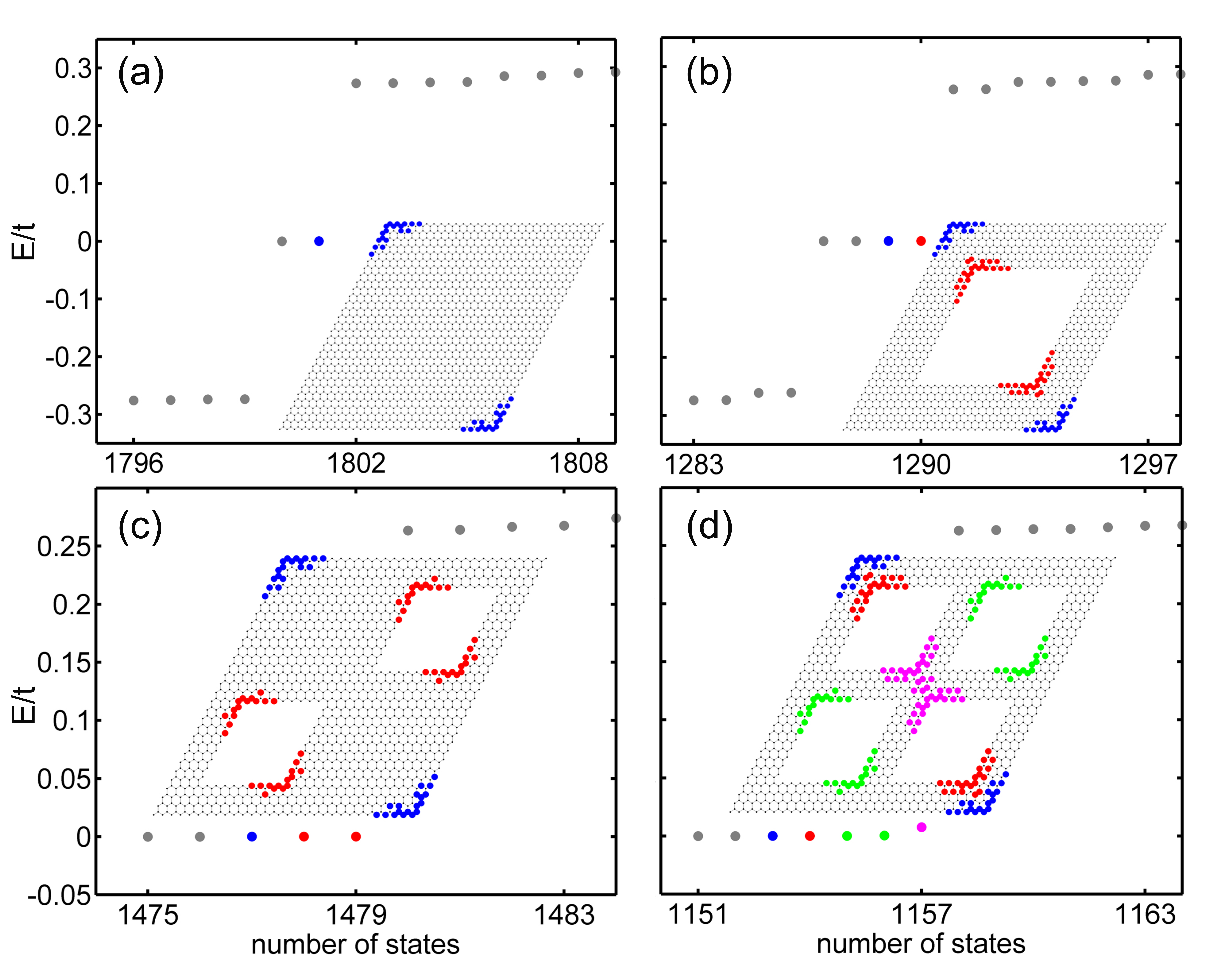}
\caption{(Color online) Discrete energy levels for the diamond-shaped flake with different hollows at $\lambda=0.3$. Eigenfunction distributions are displayed in the insets, with the same color of the corresponding energy level. (a) is for the solid flake and (b)-(d) have different hollows. }\label{fig1}
\end{figure}

In the calculation, we set $t_{SO}=0.1t$, and the Zeeman field is along y direction, i.e., $\mathbf{B}=(0,B_{y},0)$. Without loss of generality, we choose a typical diamond-shaped flake with length and width $L_x=L_y=30a$, where $a$ is the lattice constant. After diagonalizing the Hamiltonian of this isolated system, the eigenfunction distribution in Fig.\ref{fig1}(a) clearly addresses the existence of second-order topological corner states, whose energy level highlighted in blue is isolated in a large gap. Such topological corner states have fractional charge distribution, i.e., $e/2$ charge at each corner, which is independent of the Zeeman field orientation\cite{Ren2020}.

Next we introduce hollows into the diamond-shaped flake fulfilling the requirement of topological corner states. The hollow is also diamond-shaped and has obtuse interior intersections. In Fig.\ref{fig1}(b), a single hollow structure is analyzed. It is found that two pairs of zero-energy states arise in the energy gap, whose eigenfunctions are localized at corners displayed in blue and red, respectively. The interior corner states in red color are induced by this single hollow, and its energy level deviates slightly from $E=0$. Note that the energy levels of in-gap states are symmetric about $E=0$, and we only show the $E>0$ states in the following to highlight the eigenfunction distributions. When the isolated system has two identical hollows, three pairs of corner states are shown in Fig.\ref{fig1}(c), where the two red states are degenerate in energy and real spaces. For the four-hollow structure exhibited in Fig.\ref{fig1}(d), five near zero-energy states appear in the large gap, with distinguishable energy difference between each other. These numerical evidence proves that: by introducing hollows into the honeycomb lattice, multiple corner states emerge in the energy spectrum as well as real space intersections, which greatly enlarges the probability of detecting them. The multihollow structure of honeycomb lattice can be realized through fine structure engineering. For instance, the interior edges of graphene can be exactly engineered as armchair, zigzag, or hosting periodic line defects\cite{Girit,Yazyev,Acik}. Patterned graphene structures such as triangles and hexagons with zigzag boundaries have been fabricated via hydrogen-assisted anisotropic etching\cite{grapheneEngineering1,grapheneEngineering2}, which provides a solid experimental foundation for the proposed multihollow structure.

We take the four-hollow structure as the example, and introduce two transport setups for the detection and utilization of multiple corner states.

\subsection{Global resonant state in bulk insulator}

\begin{figure}[tbp]
\centering
\includegraphics[width=0.9\columnwidth]{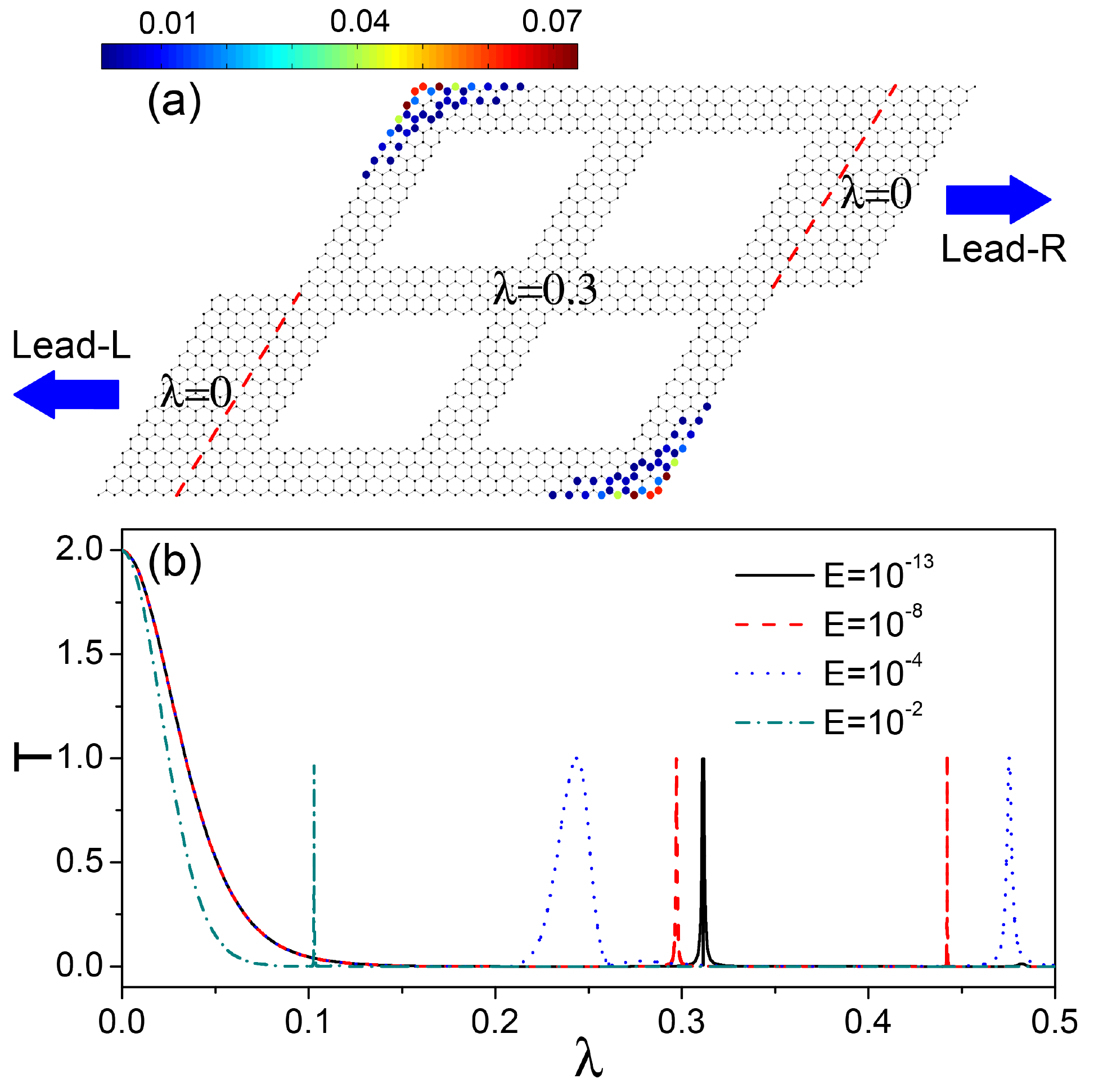}
\caption{(Color online) (a) Two-terminal transport setup, where the four-hollow structure is connected to narrow left and right leads with width $W=15a$. Red dash lines indicate the borders between regions with or without Zeeman field. (b) Transmission versus the Zeeman field strength for different energies.}\label{fig2}
\end{figure}

To study the transport features of multiple corner states, we build a two-probe system by connecting two narrow leads to the four-hollow structure, as shown in Fig.\ref{fig2}(a). In this case, all corner states remain intact in the diamond region. For instance, the eigenfunction distribution of zero-energy states locates at the exterior corners of this isolated central region. In Fig.\ref{fig2}(b), transmission spectrums with respect to the Zeeman field strength are presented. When $\lambda = 0$, quantum spin Hall edge states provide two perfect conducting channels, leading to $T=2$. In the presence of a Zeeman field, these edge states are gradually destroyed and topological corner states form, which causes the transmission quickly drops to $T=0$. With the increasing of $\lambda$, multiple $T=1$ transmission peaks emerge for energies ranging from $E=10^{-13}$ to $E=10^{-2}$. We demonstrate in the following that these peaks correspond to the resonant tunneling of corner states. Compared with the transport of one or two corner states\cite{KTWang}, the energy scope of detecting corner states is greatly enlarged in this four-hollow system.

\begin{figure}[tbp]
\centering
\includegraphics[width=\columnwidth]{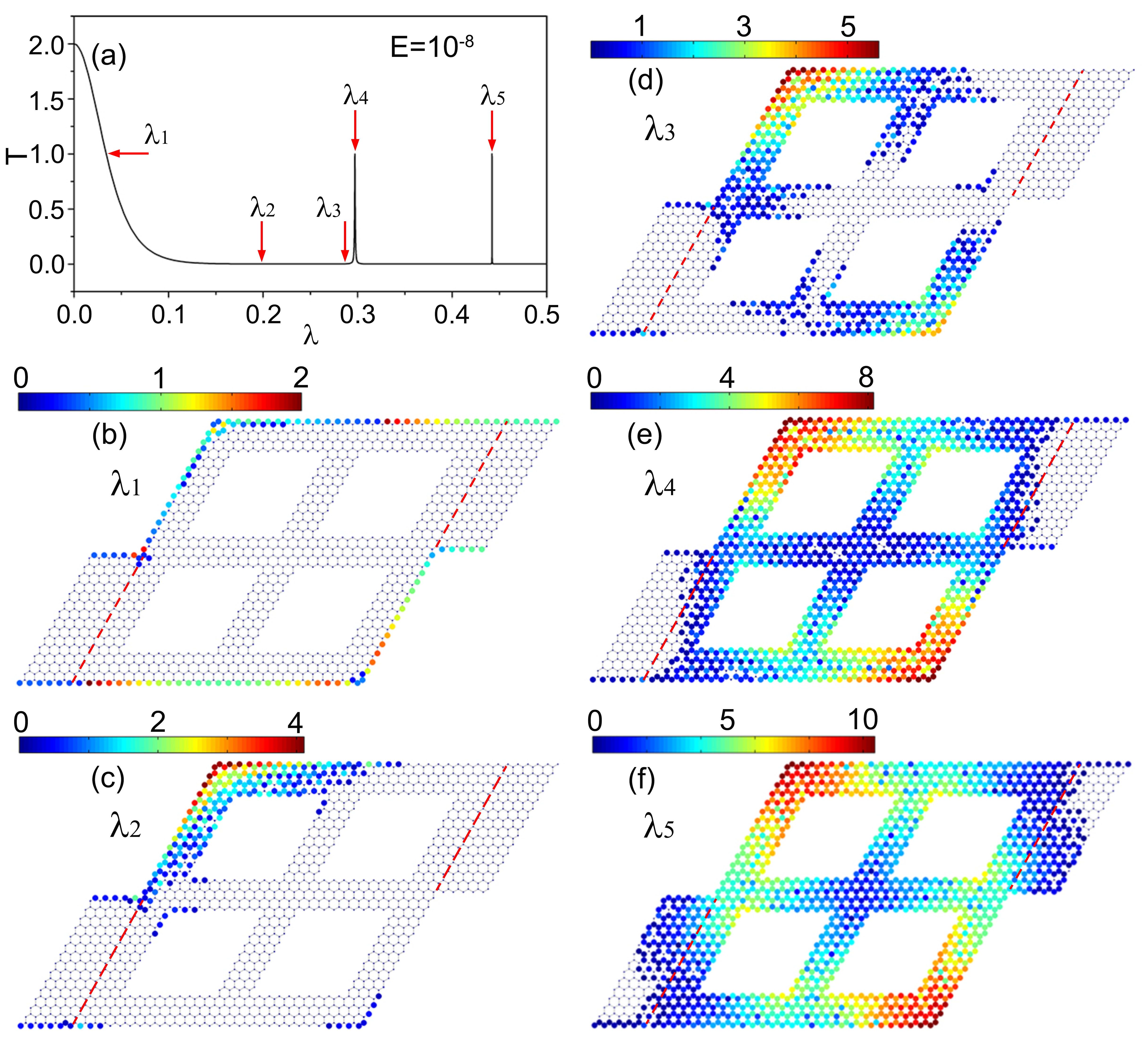}
\caption{(Color online) (a) Transmission versus Zeeman field strength $\lambda$ at $E=10^{-8}$ for the four-hollow system. Several typical $\lambda$ are labeled with red arrows. (b)-(f) Partial local density of states (PLDOS) incident from the left lead for $\lambda_1$ to $\lambda_5$. }\label{fig3}
\end{figure}

Partial local density of states (PLDOS) can effectively reveal the dynamic property of resonant tunneling process. We select one transmission curve from Fig.\ref{fig2}(b), and reproduce it in Fig.\ref{fig3}(a). As shown in red arrows, five typical Zeeman field strengths are chosen to observe their PLDOS, which are listed in Fig.\ref{fig3}(b)-(f). For a weak field $\lambda_1$, the transmission is approximately $T=1$, and Fig.\ref{fig3}(b) shows that the transport is dominated by weakened edge states. When $\lambda$ is increased to $\lambda_2$, $T \approx 0$ and corner states emerge in the PLDOS distribution. At $\lambda_3$ prior to the resonance, corner states are dressed by incident electrons and delocalize from corners. A perfect resonant peak $T=1$ occurs at $\lambda_4$, where the corresponding PLDOS occupies almost entire bulk of the system in Fig.\ref{fig3}(e). For a narrower transmission peak at $\lambda_5$, the resonant state in Fig.\ref{fig3}(f) even extends to the $\lambda=0$ region. Notice that the central region is a bulk insulator with localized corner states, and Fig.\ref{fig3}(e) and (f) show that global resonant states spread in this bulk insulating region through resonant tunneling process. Here PLDOS plays an important role in revealing the dynamic distribution of scattering states.

Similar extended states have been reported in Anderson localized systems\cite{neck1}, which are known as the necklace states\cite{neck2,neck3,neck4,neck5}. In a disordered Anderson insulator, local states trapped between potential barriers widely exist but hardly contribute to transport. As a rare event, the necklace state passes through many localized states through the multiresonance process and results in huge density of states (DOS)\cite{xu1,xu2}. Here we identify another extended state: a global resonant state in bulk topological insulator, in which incident electrons combine multiple corner states and the scattering wavefunction occupies the entire scattering region. Although originating from different mechanisms, both the necklace state and this global resonant state lead to large transmission and DOS.

\begin{figure}[tbp]
\centering
\includegraphics[width=\columnwidth]{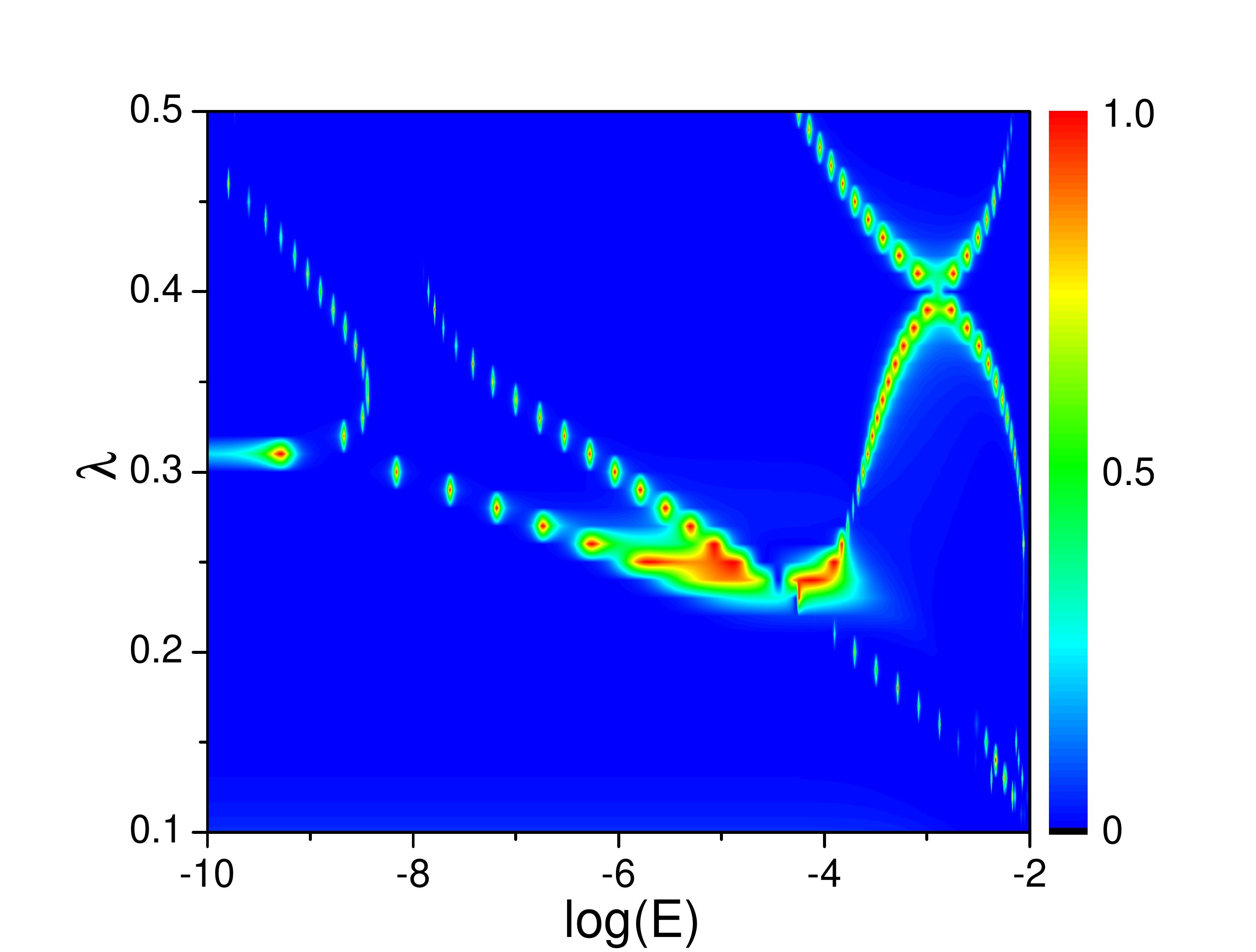}
\caption{(Color online) Transmission with respect to both Fermi energy $E$ and Zeeman field strength $\lambda$ for the two-probe system with four hollows.}\label{fig4}
\end{figure}

To summarize this subsection, we provide a 2D diagram of the transmission coefficient $T$ as a function of both Fermi energy $E$ and Zeeman field strength $\lambda$ for the two-terminal system with four hollows. By varying $E$ and $\lambda$, we see that there is a ${\cal{X}}$-shaped region with large $T$ in the upper-right corner of Fig.\ref{fig4}. Meanwhile, the bright orange region indicates that the transmission approaches $1$ under those circumstances. Besides, there are also many discrete points with high transmission, which corresponds to global resonant events in bulk topological insulator. Numerical results displayed in Fig.\ref{fig4} further confirm that, the probability of detecting topological corner states through transport measurement is greatly increased in the multihollow system.

\subsection{Single-electron source via single-parameter pumping}

\begin{figure}[tbp]
\centering
\includegraphics[width=\columnwidth]{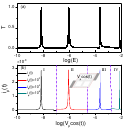}
\caption{(Color online) (a) Transmission as a function of Fermi energy for the two-probe four-hollow system at $\lambda=0.3$. (b) Instantaneous current $i_p(t)$ with respect to the pumping potential $V_p \cos (t)$ with $V_p=0.01$ in the calculation. The current spikes are scaled in magnitude for better demonstration, and labeled with Roman numerals. Inset: the single-parameter pumping setup, where a periodic gate voltage is applied on the central scattering region and accurately controls the internal potential of the two-probe system.}\label{fig5}
\end{figure}

Single-electron source is an on-demand nanoscale device in quantum computing and quantum information\cite{review2}, which reveals the dynamic aspect of coherent transport. Quantum parametric pump provides an effective way to the precise control of single-electron emitting or charge quantization\cite{niu}, which can be operated in both adiabatic and nonadiabatic regimes\cite{buttiker1,Deng,RenJ}. Recently, a silicon single-electron source with internal electron motion up to 250 GHz has been experimentally realized\cite{naturenano19}. Based on the well-separated energy levels of topological corner states in multihollow structures, we propose a single-electron source through single-parameter pumping. The system setup is shown in the inset of Fig.\ref{fig5}(b), where a periodic gate voltage is applied on the two-probe multihollow system and accurately controls its internal potential, $V(t) = V_p \cos (\omega t)$. Here $V_p$ is the amplitude of pumping potential and $\omega$ is its frequency.

We focus on the adiabatic regime. In adiabatic pumping, the internal potential changes slowly compared with the relaxation time, and the system is under equilibrium at any moment of the whole pumping period. Due to the time-dependent pumping potential $V(t)$, the instantaneous pumped current detected in the left lead is expressed as\cite{xu3}($q=1$)
\begin{equation}
i_p(t) = \frac{dQ}{dt} = \int dE (-\partial_E f ) {\rm Tr} [\Gamma_L G^{r} \frac{dV(t)}{dt} G^{a}]. \label{ip1}
\end{equation}
$f$ is the Fermi distribution function. At zero temperature, the current can be further simplified as
\begin{equation}
i_p(t) = {\rm Tr} [\Gamma_{L} G^{r}\frac{dV(t)}{dt}G^{a}], \label{ip2}
\end{equation}
where the retarded Green's function $G^{r} = ( E-H_C-V(t)-{\Sigma}_{L}^{r}-{\Sigma}_{R}^{r})^{-1}$ is time-dependent. In the calculation, we set $V_p = 0.01$ and $\omega=1$. Note that here $\omega$ is irrelevant to the energy scale and simply stands for a long pumping period $T=2\pi/\omega$. The average pumped current in one period is given by $I_p=\frac{1}{T}\int_0^T i_p(t) dt$, which is zero in this single-parameter pumping setup.

We first plot the transmission function of the two-probe multihollow system versus the Fermi energy at $\lambda=0.3$ in the absence of pumping potential. Fig.\ref{fig5}(a) shows that there are four sharp resonant peaks ($T=1$) in the transmission spectrum, which are well separated from one another. These discrete and perfect transmission peaks are ideal for single-electron manipulation. We further calculate the instantaneous pumped current $i_p(t)$ as a function of the pumping potential $V_p \cos (t)$. The Fermi energy is fixed at $E=0$ in both leads. From Fig.\ref{fig5}(b), we see that $i_p(t)$ behaves as extremely sharp spikes when $V(t)$ varies with time. It is obvious that the current spikes are closely related to the transmission peaks, where a narrower transmission peak gives rise to a higher current spike.

The discrete nature of $i_p(t)$ allows us to investigate the quantization of pumped charge. In this single-parameter adiabatic pump, the total pumped charge or current is zero in one period, since the current flowing in the central region in the first half-period will be pumped out in the second half-period. But we can integrate the current spike one by one to obtain the pumped charge, i.e., $\Delta Q_p = \int_{\Delta t} i_p(t) dt$. Numerical result gives that:
\begin{eqnarray}
\Delta Q_{p,I} = 1.0027, ~~~~\Delta Q_{p,II} = 1.0032, \nonumber \\
\Delta Q_{p,III} = 0.9969, ~\Delta Q_{p,IV} = 1.0762.
\end{eqnarray}
These quantized charges are strong evidences of single-electron pumping. Therefore, this two-probe system with multiple topological corner states can serve as a single-electron source via single-parameter pumping mechanism. Given the diagram in Fig.\ref{fig4}, we expect that this single-electron source can operate in a large range of system parameters.

\section{Conclusion}

In summary, we have investigated the quantum transport properties of a diamond-shaped honeycomb lattice with multihollow structure, where multiple second-order topological states reside in the energy spectrum as well as real-space corners. Particularly, for a system with four hollows, five pairs of corner states exist. After connecting to two narrow leads, multiple resonant peaks are found in the transmission spectrum for a relatively large energy range. Partial local density of states shows that the resonant behavior corresponds to a global resonant state, in which the scattering state expands to the whole bulk insulator. These global resonant states are induced by incident electron dressing, and reveal the dynamic transport feature of topological corner states. Based on the discrete energy levels of multiple corner states in the multihollow system, we propose a single-parameter pumping setup to realize single-electron source, and numerical results confirm that the pumped charge is perfectly quantized and well separated from one another.

\section*{acknowledgments}
This work was supported by the National Natural Science Foundation of China (Grant Nos. 12174262 and 11774238), the Natural Science Foundation of Guangdong (Grant No. 2020A1515011418), and the Natural Science Foundation of Shenzhen (Grant Nos. 20200812092737002, JCYJ20190808150409413, JCYJ20190808115415679, and JCYJ20190808152801642).

\end{document}